\documentclass[12pt]{article}
\usepackage{amsfonts}
\usepackage{epsf}
\def\be{\begin{equation}}
\def\ee{\end{equation}}
\def\bea{\begin{eqnarray}}
\def\eea{\end{eqnarray}}

\begin{document}
\begin{titlepage}
\begin{center}
{\Large \bf William I. Fine Theoretical Physics Institute \\
University of Minnesota \\}
\end{center}
\vspace{0.2in}
\begin{flushright}
FTPI-MINN-14/23 \\
UMN-TH-3348/14 \\
August 2014 \\
\end{flushright}
\vspace{0.3in}
\begin{center}
{\Large Metastable vacuum decay in center-stabilized Yang-Mills theory at large N
\\}
\vspace{0.2in}
{\bf Xin Li$^a$  and M.B. Voloshin$^{a,b,c}$  \\ }
$^a$School of Physics and Astronomy, University of Minnesota, Minneapolis, MN 55455, USA \\
$^b$William I. Fine Theoretical Physics Institute, University of
Minnesota,\\ Minneapolis, MN 55455, USA \\
$^c$Institute of Theoretical and Experimental Physics, Moscow, 117218, Russia
\\[0.2in]

\end{center}

\vspace{0.2in}

\begin{abstract}
We calculate the rate of the decay of a metastable vacuum state in $SU(N)$ gauge theory on ${\mathbb R}^3 \times {\mathbb S}^1$ space with a double trace deformation suggested by \"Unsal and Yaffe. The derived analytical expression  for the exponential factor in the decay rate, including the dependence on the parameter $\theta$, gives the exact behavior in the leading power of $N$ in the limit of large $N$ and provides a better approximation than a recently found in the literature numerical result.  
\end{abstract}
\end{titlepage}

It is well known for some time~\cite{Witten} that an $SU(N)$ gauge theory generally acquires a nontrivial nonperturbative structure of true and false vacuum states. The false vacuum states are metastable and in the limit of large $N$ are very long lived. In particular it was shown~\cite{Shifman} that in a four dimensional pure Yang-Mills theory the rate of decay of the metastable vacua is proportional to $\exp(-C N^4)$ with $C$ being an $N$ independent coefficient. More recently \"Unsal and Yaffe~\cite{uy} have suggested to consider  Yang-Mills theory formulated on the ${\mathbb R}^3 \times {\mathbb S}^1$ space with a deformation manifestly preserving the ${\mathbb Z}_N$ center symmetry. Such theory at a sufficiently small compactification size $L$ yields itself to an analytic treatment, while preserving essential features of an unmodified gauge theory, such as the confinement and a nontrivial structure of true and false vacuum states~\cite{btz}. Moreover, in the absence of an order parameter for breaking/restoration of the center symmetry the theory can be smoothly interpolated between small and large $L$ thus linking the deformed model to the unmodified one. Given this relation between the general gauge dynamics and the deformed `toy model' it is interesting to analyze the nonperturbative properties of the model in some detail. The rate of decay of a metastable vacuum in the deformed model is one such property that has been discussed most recently~\cite{btz} using a numerical calculation of the exponential factor in the rate. In the present paper we derive an analytical expression for this exponential factor, including the dependence on the parameter $\theta$. Namely we find that at large $N$ the decay rate of the lowest-energy false vacuum state to the true vacuum is given by
\be
\Gamma \sim \exp \left [ - {\cal N} \, {256 \, N^{7/2} \over 9 \, \sqrt{3} \, \pi \, (\pi - \theta)^2 } \right ]~,
\label{ff}
\ee
where ${\cal N}$ is the `semiclassicality' parameter related to the infrared scale $\Lambda$ of the Yang-Mills theory and the compactification scale $L$. At a small size $L$, such that $N \Lambda L \ll 1$, the parameter ${\cal N}$ is large~\cite{uy,btz} and a semiclassical treatment in the deformed model is justified. The formula in Eq.(\ref{ff}) is applicable at positive $\theta$,  $0 \le \theta \le 2 \pi$, and, as required on general grounds~\cite{Witten,Shifman}, is periodic in $\theta$ with the period $2 \pi$ due to interchange of the branches for the vacuum energy dependence on $\theta$. We demonstrate explicitly that the semiclassical trajectory for tunneling between the false and true vacuum, resulting in the formula (\ref{ff}), gives a lower barrier factor than the numerical results of Ref.~\cite{btz} at $N > 20$, and we argue that this trajectory in fact produces the exact  leading power behavior of the barrier factor at large $N$. 

Proceeding to derivation of Eq.(\ref{ff}) we start with briefly recapitulating the deformation of the $SU(N)$ gauge theory suggested in Ref.~\cite{uy}. The theory is formulated on the Euclidean space ${\mathbb R}^3 \times {\mathbb S}^1$, and the original Yang-Mills  action with the $\theta$ term
\be
S^{YM} =  \int_{{\mathbb R}^3 \times {\mathbb S}^1} \, d^4x \, \left \{ {1 \over 2 g^2} \, {\rm tr} \left[ F^2_{\mu \nu} (x) \right ] + {i \, \theta \over 16 \pi^2} \, {\rm tr} \left [ F_{\mu \nu} {\tilde F}^{\mu \nu} \right ] \right \}
\label{sym}
\ee
is modified by adding a polynomial in the trace of the operator for Polyakov loop wrapping around the compact dimension, $\Omega({\rm x}) \equiv {\cal P} \{ \exp [ i \oint dx_4 A_4({\rm x}, x_4)] \}$,
\be
\Delta S = \int_{{\mathbb R}^3} \, d^3 x \, {1 \over L^3} \, P \left[ \Omega({\rm x}) \right ]
\label{ds}
\ee
where $P$ is a polynomial:
\be
P [ \Omega ] = \sum_{k=1}^{[N/2]} \, a_k \, \left [ {\rm tr} \left ( \Omega^k \right ) \right ]^2
\label{po}
\ee
with $[N/2]$ standing for the integer part of $N/2$ and the positive coefficients $a_k$ can be chosen~\cite{uy} in such a way that the theory with the deformed action $S^{YM}+\Delta S$ preserves the center symmetry ${\mathbb Z}_N$. 

The compactification results in the off-diagonal $SU(N)$ gauge fields (the `$W$ bosons') developing a Kaluza-Klein mass spectrum with the lowest mass $m_W=2 \pi/NL$, and below this mass scale the dynamics is effectively Abelian, corresponding to the gauge group $U(1)^N$. The corresponding `photons' however do not stay massless, but rather acquire masses nonperturbatively due to monopoles and antimonolpoles in each of the $U(1)$ sectors.  The dynamics at the momentum scale below $m_W$ can thus be described by a dual theory in ${\mathbb R}^3$ of $N$ scalar phase fields $\sigma_n$ ($n=1,\ldots,N$)~\cite{uy}:
\be
S^{\rm dual} = { 1 \over L} \, \left ( {g \over 2 \pi } \right )^2 \, \int_{{\mathbb R}^3} \, d^3 x  \, \left [ {1 \over 2} \, \sum_{n=1}^N \, (\nabla \sigma_n)^2 - m_\gamma^2 \, \sum_{n=1}^N \, \cos \left ( \sigma_n - \sigma_{n+1} + {\theta \over N} \right ) \right ]~,
\label{sd}
\ee
where in the latter sum $\sigma_{N+1}$ is identified with $\sigma_1$, and the mass parameter $m_\gamma$ is small compared to $m_W$, provided that $N \Lambda L \ll 1$. The spectrum of the actual masses of `photons' is found by the diagonalization of the quadratic part of the dual action (\ref{sd}) and reads as
\be
m_p = m_\gamma \, \sin {\pi \, p \over N}~, ~~~~~p=1,\ldots, N-1~.
\label{mp}
\ee
It can be also noted that each of phases $\sigma_n$ is defined modulo $2\pi$ and that only the differences of the phases are dynamical, while an overall shift by a common phase $\phi$, $\sigma_n \to \sigma_n + \phi$, is not [corresponding to $p=0$ in Eq.(\ref{mp})].

Upon rescaling the coordinates ${\rm x}$ by the parameter $m_\gamma$ and adding an overall constant the action (\ref{sd}) takes the form $S= {\cal N} \, s$ with ${\cal N}= (g/2\pi)^2/m_\gamma L$ being a large `semiclassicality parameter' and $s$ standing for the `reduced' action 
\be
s= \int \, d^3 x  \, \left \{ {1 \over 2} \, \sum_{n=1}^N \, (\nabla \sigma_n)^2 +  \sum_{n=1}^N \, \left [ 1 - \cos \left ( \sigma_n - \sigma_{n+1} + {\theta \over N} \right ) \right ]  \right \}~.
\label{sdn}
\ee 
In what follows we work with this reduced action, so that the relevant quantities, e.g. the energy and the barrier factor, are calculated `in units of ${\cal N}$\,'.

The potential in Eq.(\ref{sdn}) develops local minima. The two lowest of these at $0 \le \theta \le 2\pi$ are~\footnote{Due to the $2 \pi$ periodicity in $\theta$ it is sufficient to consider only one period. We also do not consider the higher metastable states corresponding (in the first period in theta) to $\sigma_n = 2 \pi \, k \, n/N$ with $k > 1$.}
\be
| {\rm vac1} \rangle: ~~~ \sigma_n =0~~~~{\rm and}~~~~| {\rm vac2} \rangle: ~~~\sigma_n = {2 \, \pi \, n \over N}~.
\label{v12}
\ee
and the corresponding values of the energy are given by
\be
E_1 = N \, \left ( 1-\cos {\theta \over N} \right ) \to {\theta^2 \over 2 N}, ~~~~~ E_2 = N \, \left ( 1- \cos{2 \pi - \theta \over N } \right ) \to {(2 \pi - \theta)^2 \over 2 N}~,
\label{e12}
\ee
where the limiting expressions are written for large $N$. Clearly, these two states become degenerate and intersect at $\theta=\pi$ where the difference between their energy changes sign: 
\be
\epsilon = E_2 - E_1 \to {2 \pi \, (\pi - \theta) \over N}~.
\label{eps}
\ee

Depending on the value of $\theta \neq \pi$, the lower of the vacuum states is stable while the other one is metastable and decays to the lower one by tunneling, as discussed in the context of this model in Ref.~\cite{btz}. In what follows we calculate the exponential factor in the tunneling rate for this process. We start with discussing the case where $\theta=0$ (so that $E_2 > E_1$) and then generalize the result to nonzero values of $\theta$. The tunneling in the three dimensional model is described~\cite{coleman,cc} by a nontrivial $O(3)$ symmetric solution to the classical equations of motion approaching $|{\rm vac2} \rangle$ at infinity (and approaching vicinity of $|{\rm vac1} \rangle$ at the center), which solution is called `bounce'. The tunneling rate $\Gamma$ per  three dimensional volume $V$ is then given by $\Gamma/V \sim \exp(-S_B)$, where $S_B$ is the action for the bounce. At small energy difference $\epsilon$ the bounce can be considered and its action calculated in the so-called `thin wall approximation'\cite{kov,coleman}. Namely, in the limit $\epsilon \to 0$, the two vacua are degenerate and there is a one dimensional classical configuration interpolating between $|{\rm vac2} \rangle$ at $x \to -\infty$ and $|{\rm vac1} \rangle$ at $x \to +\infty$ with a one-dimensional action $S_1$ (the surface tension of the wall).  In three dimensions, when the wall is wrapped into a sphere of the radius $R$ the action associated with the surface of the sphere can be taken as $4 \pi \,  R^2 \, S_1$, provided that the radius is much larger than any scale for the thickness of the transition region in the one dimensional solution. Restoring a nonzero energy difference $\epsilon$ between the exterior and the interior of the sphere one can write the action for the configuration with a spherical wall as
\be
S(R)= 4 \pi \,  R^2 \, S_1 - {4 \pi \over 3} \, R^3 \, \epsilon~,
\label{sr}
\ee
where the last term describes the contribution to the action of the energy difference in the interior volume of the sphere. The extremum of the action $S(R)$ defines the bounce radius $R_B=2 \, S_1/\epsilon$ and the bounce action 
\be
S_B={16 \pi \over 3} \, {S_1^3 \over \epsilon^2}~.
\label{sbr}
\ee
One can notice, in particular, that the thin wall approximation is always justified at small $\epsilon$.

In the problem at hand the energy difference $\epsilon$ is given by Eq.(\ref{eps}) so that it remains to calculate the surface tension, i.e. the action for a one dimensional configuration interpolating between the two vacua in the limit where they are degenerate in energy. Unlike a calculation for a  true multidimensional bounce, which does not correspond to a minimum of the action [as can be seen e.g. from Eq.(\ref{sbr})], the wall interpolating between two degenerate vacua does correspond to the minimum of the one dimensional action for the given boundary conditions at $x \to \pm\infty$. For this reason in the calculation of $S_1$ we start with a trial configuration and then argue that it provides the leading in $N$ behavior of $S_1$. In order to describe this trial configuration we introduce `the center of gravity' $\Sigma$ for the phases $\sigma_n$ and the `distance' $\sigma$ between the last and the first phases:
\be
\Sigma = {1 \over N} \sum_{n=1}^N \sigma_n,~~~~~\sigma=\sigma_N - \sigma_1~.
\label{sdef}
\ee
In the interpolating configuration that we consider the phases $\sigma_n$ are equally spaced in $n$ between $\sigma_1$ and $\sigma_N$:
\be
\sigma_n = \Sigma + \left ( {n-1 \over N-1} - {1 \over 2} \right ) \, \sigma \to \left ( {n \over N} - {1 \over 2} \right ) \, \sigma~,
\label{str}
\ee
where in the last transition we have set the limit of large $N$ and also set the overall phase $\Sigma$ to zero, since this common shift, as previously mentioned, is not dynamical. 

One readily finds that the reduced action from Eq.(\ref{sr}) in this configuration depends only on the variable $\sigma$,   whose dependence on $\sigma(x)$, and this dependence at large $N$ (and $\theta=0$) is given by 
\be
s= \int \, d^3x \, \left [ {N \over 24} \, (\nabla \sigma)^2 + 1 - \cos \sigma + {\sigma^2 \over 2 N} \right ]~.
\label{srtr}
\ee
The last term in the integrand corresponds to the energy difference between the local minimum at $\sigma = 2 \pi$ and the global at $\sigma=0$ and, according to the previous discussion describes the parameter $\epsilon$, while the rest of the action is that of the Sine-Gordon model, so that the interpolating trajectory for $\sigma(x)$ between the two minima is the well known kink profile:
\be
\sigma(x) = 4 \arctan \left [ \exp \left ( - 2 \, \sqrt{3 \over N} \, x \right ) \right ]~,
\label{sgp}
\ee
whose action is given by 
\be
s_1= 4 \, \sqrt{N \over 3}
\label{s1sg}
\ee

One can notice that an introduction of a nonzero value of $\theta$ with the described trial configuration modifies only the energy difference $\epsilon$ between the two vacua according to Eq.(\ref{eps}), but does not affect the Sine-Gordon part of the action in Eq.(\ref{srtr}), i.e. it does not change the expression (\ref{s1sg}). Thus one can use the equations (\ref{eps}), (\ref{sbr}) and (\ref{s1sg}) to readily arrive at the formula in Eq.(\ref{ff}). It should be also noted that the bounce radius in this solution scales as $R_B \sim N^{3/2}$ and is thus much larger than the distance scale $O(N^{1/2})$ inherent in the profile (\ref{sgp}) as well as the scale $O(N)$ corresponding to the lowest mass in Eq.(\ref{mp}). This conclusion about applicability of the thin wall approximation is qualitatively similar to the one derived in Ref.~\cite{btz}.

\begin{figure}[ht]
\begin{center}
 \leavevmode
    \epsfxsize=12cm
    \epsfbox{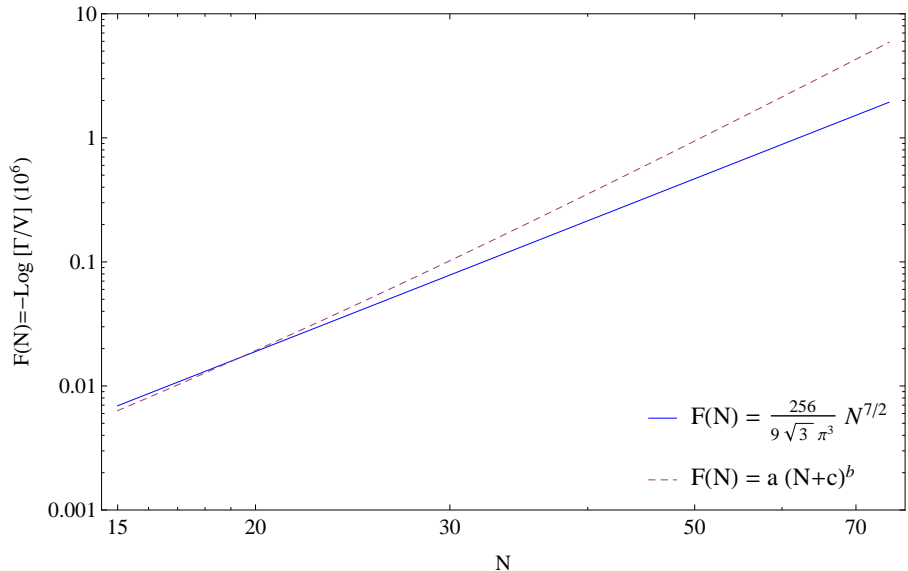}
    \caption{Comparison of the barrier factor $F(N)= -\log (\Gamma/V)$ (in units of $10^6$) given by Eq.(\ref{ff}) at $\theta=0$ (solid) with the fit to the numerical data of Ref.~\cite{btz}, $F(N)= a \, (N+c)^b$ with $a=3.906 \times 10^{-3},~ b=4.83304,~ c=4.26324$ (dashed).}
\end{center}
\end{figure} 

We show in Fig.~1 the comparison of the exponential factor (at $\theta=0$) described by Eq.(\ref{ff}) with the interpolation given in Ref.~\cite{btz} for their numerical results. It is quite clear from this comparison that our trial trajectory gives a better approximation starting from $N \approx 20$. We believe that one possible reason for the deviation in the numerical calculation of Ref.~\cite{btz} is the adopted there choice of the size of the `box' in $x$. The criterion for sufficiently `infinite' cutoff in $x$ used there was based on the value of the exponential factor $\exp(-x)$, and the interval $|x| \le 16$ apparently was used. In our trial trajectory the solution (\ref{sgp}) for the Sine-Gordon problem involves a distance scale of order $\sqrt{N}$ rather than of order one, so that evaluating the effect of the boundary conditions at finite $x$ from the $\exp(-x)$ may not be quite correct at large $N$. Furthermore, if the lowest `photon' modes, corresponding to $p= O(1)$ in Eq.(\ref{mp}) are present in the numerical simulation, this may introduce a distance scale in the calculation as long as of order $N$. 

Clearly, the actual action $s_1$ should be smaller than that in Eq.(\ref{s1sg}) found on the trial configuration. It can be argued however, that the difference at $N \to \infty$ is not larger than of order one, or less, so that Eq.(\ref{s1sg}) gives the leading asymptotic behavior. Indeed, the exact trajectory generally differs from that described in Eq.(\ref{str}) by a different from equidistant distribution of the intermediate values of $\sigma_n(x)$ between $\sigma_1(x) = -\sigma(x)/2$ and $\sigma_N = \sigma(x)/2$: $\sigma_n(x)=f(n,\sigma)$. At large $N$ the difference between successive variables $\sigma_n - \sigma_{n+1}$ is  small, and one can replace the discrete summation by integration over $n$, thus rewriting the one dimensional (in $x$) action with degenerate vacua in the form
\be
s=\int_{-\infty}^\infty d x \, \int_0^N dn \ \left [ {1 \over 2} \, \left ( {\partial f \over \partial x} \right )^2 + {1 \over 2} \, \left ( {\partial f \over \partial n} \right )^2 \right ] + \int dx \, \left ( 1- \cos \sigma - {\sigma^2 \over 2 N} \right )~,
\label{seqv}
\ee
with the boundary conditions $f(N,x)=-f(0,x)=\sigma(x)/2$, and $\sigma(x)$ yet to be determined from minimization of the action. (We also have extended, for simplicity of notation, the counting of $n$ from $1 \le n \le N$ to $0 \le n \le N$ which does not affect the leading $N$ behavior.) A general configuration can be written in a form of a perturbation over the trial trajectory: 
\be
f(n,x) =  \left ( {n \over N} - {1 \over 2} \right ) \, \sigma(x) + h(n,x)~,
\label{f0}
\ee
with zero boundary conditions, $h(0,x)=h(N,x)=0$ and $h \to 0$ at $x \to \pm \infty$, and the action takes the form
\bea
&& s=  \int \, dx \, \left [ {N \over 24} \, ( \sigma')^2 + 1 - \cos \sigma \right ] + \nonumber \\
&&\int_{-\infty}^\infty d x \, \int_0^N dn \ \left [ {1 \over 2} \, \left ( {\partial h \over \partial x} \right )^2 + {1 \over 2} \, \left ( {\partial h \over \partial n} \right )^2  - \left ( {n \over N} - {1 \over 2} \right ) \, \sigma''(x) \, h \right ]~,
\label{sh}
\eea
where $\sigma'(x) \equiv d \sigma/dx$ and $\sigma''(x) \equiv d^2 \sigma/dx^2$. One can notice that the term $\sigma^2/2N$ in the last integrand in Eq.(\ref{seqv}) is canceled at $x \to \pm \infty$ by the contribution of the linear in $n$ term from Eq.(\ref{f0}) to the first integral, so that the vacua corresponding to the action (\ref{sh}) are degenerate. The discussed above trial configuration corresponds to identically vanishing $h$. Clearly, the Eq.(\ref{sh}) contains a source of $h$ proportional to $\sigma''(x)$ so that in the exact solution $h(n,x)$ is not vanishing and the action receives an extra negative contribution. One can however notice that the distance scale in $\sigma''(x)$ found on the trial trajectory is of order $N^{1/2}$, so that it is at such distances both in $x$ and $n$ from the transition region at $(n,x) = (0,0)$ and $(n,x)=(N,0)$ that a nonzero $h$ is generated. 
Therefore, given that in the transition region $\sigma'' \sim O(N^{-1})$ and the `area' of the region is $\Delta x \, \Delta n \sim O(N)$ it can be concluded that  the contribution of the second integral term in Eq.(\ref{sh}) can be estimated as being of order one, while the first integral term is of order $N^{1/2}$, as described by Eq.(\ref{s1sg}). Thus the deviation of the result in the latter equation from the exact expression is sub leading at large $N$.~\footnote{Furthermore, the deviation can be additionally suppressed at large $N$ by cancellations due to the antisymmetry in $x$ and $n$ of the source term. However, this possibility is beyond our present approximate treatment.}

In conclusion. Among the theoretically tractable interesting nonperturbative features of a center-stabilized deformed $SU(N)$ gauge theory on the ${\mathbb R}^3 \times {\mathbb S}^1$ space at a small compactification length~\cite{uy} is the problem of tunneling between its metastable and stable vacuum states~\cite{btz} that can be analyzed in terms of the dual formulation with the action in Eq.(\ref{sd}). We have considered a trial configuration describing such tunneling and resulting in the analytical expression in Eq.(\ref{ff}) for the exponential factor in the false vacuum decay rate. We have also argued that this expression in fact gives the leading behavior at large $N$.

The work of M.~B.~V.  is supported, in part, by the DOE grant DE-SC0011842.

\end{document}